\documentclass[aps,prb,superscriptaddress,floatfix,10pt,twocolumn]{revtex4-1}
\usepackage{graphicx,graphics}
\usepackage{dcolumn}
\usepackage{amsmath,amssymb,amsfonts}
\usepackage{latexsym,verbatim}
\usepackage{bm}
\usepackage{color}
\usepackage{ulem}
\usepackage{multirow}
\usepackage[abs]{overpic}
\usepackage[breaklinks=true,colorlinks,citecolor=blue,linkcolor=blue,urlcolor=blue]{hyperref}
\usepackage{overpic}
\usepackage{array}

\begin{document}
\title{Electronic structure and magnetic properties of few-layer Cr$_2$Ge$_2$Te$_6$: the key role of nonlocal electron-electron interaction effects}
\author{Guido Menichetti}
\email{guido.menichetti@iit.it}
\affiliation{Istituto Italiano di Tecnologia, Graphene Labs, Via Morego 30, I-16163 Genova,~Italy}

\author{Matteo Calandra}
\affiliation{Sorbonne Universit\'e, CNRS, Institut des Nanosciences de Paris, UMR7588, F-75252, Paris, France}

\author{Marco Polini}
\affiliation{Istituto Italiano di Tecnologia, Graphene Labs, Via Morego 30, I-16163 Genova,~Italy}

\begin{abstract}
Atomically-thin magnetic crystals have been recently isolated experimentally, greatly expanding the family of two-dimensional materials. 
In this Article we present an extensive comparative analysis of the electronic and magnetic properties of ${\rm Cr}_2{\rm Ge}_2{\rm Te}_6$,
based on density functional theory (DFT). 
We first show that the often-used ${\rm DFT}+U$ approaches fail in predicting the ground-state properties of this material in both its monolayer and bilayer forms, and even more spectacularly in its bulk
form. In the latter case, the fundamental gap {\it decreases} by increasing the Hubbard-$U$ parameter, eventually leading to  a metallic ground state for physically relevant values of $U$, in stark contrast with experimental data. On the contrary, the use of hybrid functionals, which naturally take into account nonlocal exchange interactions between all orbitals, yields good account of the available ARPES experimental data. We then calculate all the relevant exchange couplings (and the magneto-crystalline 
anisotropy energy)  for monolayer, bilayer, and bulk ${\rm Cr}_2{\rm Ge}_2{\rm Te}_6$ with a hybrid functional, with super-cells containing up to $270$ atoms, commenting on existing calculations with much smaller super-cell sizes. In the case of bilayer ${\rm Cr}_2{\rm Ge}_2{\rm Te}_6$, we show that two distinct intra-layer second-neighbor exchange couplings emerge, a result which, to the best of our knowledge, has not been noticed in the literature.
\end{abstract}

\maketitle

\section{Introduction}
\label{sect:intro}

The family of two-dimensional (2D) materials beyond graphene is now very large~\cite{Geim2013,Mounet2018} and contains also magnetic crystals~\cite{Shabbir2018,Burch2018,Gong2019}, 
notably ${\rm Cr}_2{\rm Ge}_2{\rm Te}_6$~\cite{Gong2017,Wang2018,Kim2019}, ${\rm Cr}{\rm I}_3$~\cite{Huang2017,Wang2018a,Klein2018,Huang2018,Song2018,Jiang2018a,Kim2018,Jin2018,Jiang2018b,Song2019,Thiel2019}, ${\rm V}{\rm Se}_2$~\cite{Bonilla2018}, ${\rm Fe}_3{\rm Ge}{\rm Te}_2$~\cite{Fei2018,Tan2018,Deng2018,Wang2019}, ${\rm Fe}{\rm P}{\rm S}_3$~\cite{Wang2016b,Lee2016}, ${\rm Mn}{\rm Se}_2$~\cite{Ohara2018}, ${\rm Cr}{\rm Br}_3$~\cite{Kim2019,Ghazaryan2018,Zhang2019}, ${\rm Cr}{\rm Cl}_3$~\cite{Klein2019}, and ${\rm Gd}{\rm Te}_3$~\cite{Lei2019b}.

The atomic nature of these materials is obviously rather complex and a microscopic understanding of their ground-state properties requires massive use of {\it ab initio} theories, especially density-functional theory (DFT)~\cite{Giuliani2005}. In the context of few-layer magnetic crystals, the greatest deal of efforts has
been devoted to the study of ${\rm CrI}_3$~\cite{Wang2018a,Zhang2015,Soriano2018,Webster2018,Webster2018a,Jiang2018,Zheng2018,McGuire2015,Wang2011,Liu2016,Lado2017,Wang2016a,Gudelli2019,Xu2018,Lee2019,Sivadas2018,Jang2018,Djurdji2018,Abramchuk2018,Liu2019,Torelli2018,Lei2019},
while ${\rm Cr}_2{\rm Ge}_2{\rm Te}_6$ appears to be less studied~\cite{Gong2017,Wang2018,Xu2018,Li2014b,Sivadas2015,Zhuang2015,Li2018,Fang2018,Kang2019,Dong2019,Wang2019b,Sun2019} and rich in controversial findings, as we discuss below.

In this work we therefore focus our attention on ${\rm Cr}_2{\rm Ge}_2{\rm Te}_6$. Bulk ${\rm Cr}_2{\rm Ge}_2{\rm Te}_6$ is a ferromagnetic semiconductor~\cite{Carteaux1995,Siberchicot1996} below a Curie temperature $T_{\rm C} \sim 60~{\rm K}$, with an out-of-plane magnetic 
easy axis and negligible coercivity. Scanning magneto-optical Kerr microscopy has been carried out~\cite{Gong2017} in bilayer ${\rm Cr}_2{\rm Ge}_2{\rm Te}_6$, 
revealing ferromagnetic order below $T_{\rm C} \sim 30~{\rm K}$. Electric-field control of the magnetization 
direction has been demonstrated~\cite{Wang2018} in field-effect transistors made of ${\rm Cr}_2{\rm Ge}_2{\rm Te}_6$, with variable thicknesses in the range $3.5$-$20~{\rm nm}$. 
Ballistic Hall micromagnetometry~\cite{Geim1997,novoselov2003} has finally been applied to six-layer ${\rm Cr}_2{\rm Ge}_2{\rm Te}_6$ to study the temperature 
dependence of the magnetization---see Supplementary Figure S4 of Ref.~\onlinecite{Kim2019}.

Widely used approximations for the exchange and correlation (xc) energy functional in DFT,  mainly  based on parametrizations of the xc energy of the homogeneous electron gas~\cite{Giuliani2005}, 
miss  important features  of the  physical properties of these magnetic materials.  For  instance,  both  the local spin-density approximation  (LSDA) and the spin-polarized generalized gradient approximation (spGGA) fail  in describing the insulating behavior of systems with strongly localized and thus, correlated,  valence electrons. Leaving aside the well-known ``gap problem"~\cite{Giuliani2005}, these approximations often produce a qualitatively wrong ground state.
Due to their moderate computational cost and  the ability to capture  correlation  effects, the  Hubbard-corrected  LSDA (``${\rm LDA}+U$'') and spGGA (``${\rm GGA}+U$'') approximations~\cite{anisimov_prb_1991,anisimov_prb_1993,solovyev_prb_1994,Liechtenstein1995} have been extensively used. 

Many authors~\cite{Gong2017,Wang2018,Xu2018,Li2014b,Sivadas2015,Li2018,Fang2018,Kang2019,Dong2019,Sun2019} studied the electronic and magnetic properties of  ${\rm Cr}_2{\rm Ge}_2{\rm Te}_6$ using the
${\rm LDA}+U$/${\rm GGA}+U$ approximations in order to deal with electrons belonging to the semi-filled $3d$ orbitals of 
Cr. The main objection we raise to these approaches is that strong hybridization between Cr $d$ and Te $p$ orbitals pushes ${\rm Cr}_2{\rm Ge}_2{\rm Te}_6$ away from the atomic limit, invalidating the use of Hubbard-type descriptions.  
Moreover, we recall that arbitrary choices of the Hubbard-$U$ parameter have been criticized in the literature~\cite{Cococcioni2005,Campo2010,Kulik2006,Kulik2010}, where it has been shown that $U$ is an intrinsic response property which should be calculated self-consistently. 

A few calculations~\cite{Li2014b,Zhuang2015,Fang2018,Kang2019,Wang2019b} have also been carried out using hybrid functionals, which mix
the local (or semi-local) DFT exchange with the exact nonlocal Hartree-Fock (HF) exchange, 
to treat inaccuracies of the LDA/GGA approximations. However, none of these studies compared results obtained by using different approaches, highlighting the criticality of the ${\rm LDA}+U$/${\rm GGA}+U$ approximations for this specific magnetic material.

In this Article we instead present a DFT study of the ground state properties of ${\rm Cr}_2{\rm Ge}_2{\rm Te}_6$, comparing the results obtained in the ${\rm GGA}+U$ approximation (in both its non-self-consistent and self-consistent versions) with those obtained via the use of a hybrid functional.

In order to understand the microscopic properties 
of magnetic materials it is crucial to derive effective spin models---such as the Heisenberg Hamiltonian~\cite{Gubanov1992, Auerbach1994}---from
{\it ab initio} calculations. Lichtenstein et al.~\cite{Liechtenstein1987} (LKAG) introduced a rigorous approach to evaluate the exchange couplings of the Heisenberg model from {\it ab initio} calculations. These parameters strongly depend on the local environment around a magnetic ion, such as the spin configuration itself, especially
in metals and small-gap semiconductors where the spin density is weakly inhomogeneous. In the LKAG approach~\cite{Liechtenstein1987}, the exchange parameters are determined by studying the total energy variation with respect to small deviations of the magnetic moments of interest from the ground-state magnetic configuration. It relies on the application of the Andersen's ``local force theorems''~\cite{Machintosh1980,Methfessel1982}, which ensure its
validity even in the aforementioned complex cases.

Applying the theory of Ref.~\onlinecite{Liechtenstein1987} in full glory is however computationally demanding and therefore approximate approaches are often used. (For recent work utilizing the LKAG approach in combination with the LSDA and spGGA on bulk ${\rm Cr}{\rm Cl}_3$ and ${\rm Cr}{\rm I}_3$ see e.g.~Ref.~\onlinecite{Besbes2019}.) These need to be handled with great care. In particular, the exchange coupling parameters
and magneto-crystalline anisotropy of few-layer and bulk ${\rm Cr}_2{\rm Ge}_2{\rm Te}_6$ have been calculated by DFT methods using two of such approximate methods:

i) One, which will be dubbed below ``FM-AFM" approach, relies on calculating the difference between the total energies of the ferromagnetic (FM) and anti-ferromagnetic (AFM) configurations. The FM-AFM approach was used by the authors of Refs.~\onlinecite{Li2014b,Zhuang2015,Dong2019} to calculate the intra-layer nearest-neighbor exchange couplings. Similarly, the authors of Refs.~\onlinecite{Sivadas2015,Fang2018} used the same approach but calculated exchange couplings up to third-neighbors. The fact that such couplings are important is deduced from inelastic neutron scattering measurements of spin waves in ${\rm MnPS}_3$~\cite{Wildes1998} and ${\rm FePS}_3$~\cite{Wildes2012}. The authors of Ref.~\onlinecite{Fang2018} evaluated also the nearest-neighbor inter-layer
exchange coupling in few-layer and bulk ${\rm Cr}_2{\rm Ge}_2{\rm Te}_6$. 

The evaluation of the exchange couplings via the FM-AFM approach is not justified~\cite{Liechtenstein1987} when the ground-state spin density is weakly inhomogeneous and the dependence of the exchange couplings on the magnetic configuration is strong, as in the case of metals and small-gap semiconductors.

ii) Another method, dubbed below ``four-state mapping analysis'' (FSMA)~\cite{Whangbo2003,Xiang2011,Xiang2013a}, also relies on total-energy calculations but focuses only on the spin configurations of the two ions ``linked'' by the exchange coupling of interest, leaving all the other ions in the ground-state configuration. The FSMA approach has been used for ${\rm Cr}_2{\rm Ge}_2{\rm Te}_6$ by the authors of Refs.~\onlinecite{Gong2017,Li2018,Xu2018}. 
Xu et al.~\cite{Xu2018} used the FSMA approach to study the intra-layer 
exchange couplings between nearest-neighbor Cr atoms in monolayer ${\rm Cr}_2{\rm Ge}_2{\rm Te}_6$. The same method has been used by the authors of Refs.~\onlinecite{Gong2017,Li2018} to evaluate the intra- and inter-layer exchange 
couplings up to third-neighbors for bulk ${\rm Cr}_2{\rm Ge}_2{\rm Te}_6$. While the FSMA method is a good compromise between the rigor of the LKAG approach and computational feasibility, it is very sensitive to the super-cell size. In Ref.~\onlinecite{Xu2018} a $2\times 2$ super-cell was used, while in Refs.~\onlinecite{Gong2017,Li2018} super-cells up to dimension $2\sqrt{3} \times 2 \times 1$ were used. Below,
we demonstrate that such super-cell dimensions do not give converged values of the exchange coupling parameters. 

Finally, we show that bilayer ${\rm Cr}_2{\rm Ge}_2{\rm Te}_6$ displays two inequivalent ${\rm Cr}$ atoms, a fact that leads to the existence of two distinct intra-layer second-neighbor exchange couplings, $J^{\rm a}_{2}$ and $J^{\rm b}_{2}$. To the best of our knowledge, this has not been noticed in the literature.

This Article is organized as following. In Sect.~\ref{sect:theory} we present the geometrical structure of ${\rm Cr}_2{\rm Ge}_2{\rm Te}_6$ (Sect.~\ref{sect:geometry}) and the theoretical methods (Sect.~\ref{sect:technical_details}) that we have used to compute the electronic properties of this material both in the few-layer and bulk cases. In Sect.~\ref{sect:numerical_results} we present a summary of our main results on the electronic structure of few-layer and bulk ${\rm Cr}_2{\rm Ge}_2{\rm Te}_6$, comparing the ${\rm GGA}+U$ approach with the hybrid-functional one. Exchange couplings and magneto-crystalline anisotropy are reported in Sect.~\ref{sect:Heisenberg}. A summary of our main findings and a brief set of conclusions are finally presented in Sect.~\ref{sect:conclusions}.

\section{Theoretical approach}
\label{sect:theory}

In this Article, we study the electronic and magnetic properties of the transition metal trichalcogenide ${\rm Cr}_2{\rm Ge}_2{\rm Te}_6$ in its monolayer, bilayer, and bulk forms. 

In Sect.~\ref{sect:technical_details}, which can be skipped by uninterested readers, we discuss the computational tools we have used. In Sect.~\ref{sect:geometry} we present geometric and structural details.

\subsection{Computational details}
\label{sect:technical_details}
We carry out DFT calculations by using the \textsc{Quantum Espresso} (\textsc{QE})~\cite{QE1,QE2} and \textsc{CRYSTAL14}~\cite{CRYSTAL14} codes, which use plane waves and atom-centered (Gaussian) basis sets, respectively. 
    
For the calculations with \textsc{QE}  we use the SG15 Optimized Norm-Conserving Vanderbilt (ONCV) pseudopotentials~\cite{Hamann2013,Schlipf2015,Scherpelz2016}, generated by taking into account both the scalar- and fully-relativistic potentials, in order to turn off/on the contribution from  spin-orbit coupling (SOC). The pseudopotentials include also the semi-core states and their 
 valence configurations are: $3s^2,\ 3p^6,\ 4s^2,\ 3d^4$ for ${\rm Cr}$; $4d^{10},\ 5s^2,\ 5p^4$ for ${\rm Te}$; and
$3d^{10},\ 4s^2,\ 4p^2$ for ${\rm Ge}$. We use an energy cutoff up to $70~{\rm Ry}$ for all the calculations. 
For the BZ integrations we employed a Marzari-Vanderbilt smearing~\cite{Marzari1999} of $10^{-4}~{\rm Ry}$ with a Monkhorst-Pack (MP)~\cite{MP} ${\bm k}$-point grid with $6\times 6\times 6$~($12\times 12\times 12$)  points for self-consistent calculations of the charge density (density of states) for the bulk crystal, while for the few-layer crystals we adopt a grid with $8\times8$~($16\times16$) points for self-consistent calculations of the charge density (density of states).
For the calculations with \textsc{CRYSTAL14} we used a 86-411d41 Gaussian all-electron basis set~\cite{Catti1996} with $24$ valence electrons for ${\rm Cr}$ and a double-zeta basis set with an effective core pseudo-potential 
(m-cc-pVDZ-PP)~\cite{Heyd2005} with $24$ valence electrons for ${\rm Te}$ and $22$ valence electrons for ${\rm Ge}$. The charge density integrations over the BZ are performed using a uniform  
$12\times 12\times 12$ ($12\times12$) MP ${\bm k}$-point grid for bulk (few-layer) ${\rm Cr}_2{\rm Ge}_2{\rm Te}_6$, respectively. When we use a super-cell, we scale the  MP grid size to assure the same accuracy as in the single-cell calculations.

The xc potential is treated in the ${\rm spGGA}$, as parametrized by the Perdew-Burke-Ernzerhof (PBE) formula~\cite{PBE}, with the van der Waals (vdW)-D2 correction
proposed by Grimme~\cite{Grimme2006}. In order to improve the description of the semi-filled  $3d$ orbitals of the ${\rm Cr}$ atoms, we adopt both the ${\rm GGA}+U$  scheme~\cite{Cococcioni2005,Kulik2006,Kulik2010} 
and the HSE06~\cite{HSE06} hybrid xc functional, which mixes the PBE exchange with $25\%$ of the exact nonlocal HF exchange. This allows us to compare results obtained with different approaches, pointing out an evident criticality of the ${\rm GGA}+U$ method for the material of interest, as discussed below.
As anticipated in Sect.~\ref{sect:intro}, the Hubbard-$U$ parameter is an intrinsic response property of the material. We calculate it self-consistently in the spin-polarized case, following the linear response approach of Refs.~\onlinecite{Cococcioni2005,Kulik2006,Kulik2010}. For bulk ${\rm Cr}_2{\rm Ge}_2{\rm Te}_6$ we find a self-consistent value of $U$ given by $U_{\rm sc}\equiv 3.9~{\rm eV}$. We use this value for all the few-layer crystals under investigation (i.e.~monolayer and bilayer ${\rm Cr}_2{\rm Ge}_2{\rm Te}_6$) because we have checked that $U_{\rm sc}$ changes slightly with the number of layers.
We also employ $U=U_{\rm sc}$ for the ${\rm GGA}+U$ non-collinear calculations, which include SOC. Finally, we also study the electronic properties, with and without SOC, by varying the value of the Hubbard-$U$ parameter in the range of $0<U\leq U_{\rm sc}=3.9$, as explained below.

In order to evaluate fully-relativistic electronic band structures with the HSE06 hybrid functional we used the \textsc{Wannier90}~\cite{W90} code. In fact, it is not possible to calculate them directly from \textsc{QE} due to the heavy computational cost, and SOC is not yet implemented in the \textsc{CRYSTAL14} code. The \textsc{Wannier90} code allows us to overcome this problem interpolating the electronic band structure using Maximally-Localized Wannier Functions (MLWFs)~\cite{Marzari1997,Souza2001}  extracted from the DFT calculations with \textsc{QE}. As starting guess for the ``Wannierisation" procedure, we project the Bloch states onto trial localised atomic-like orbitals: $d$-orbitals for the ${\rm Cr}$ atoms, $p$-orbitals for ${\rm Te}$  atoms, and $p_z$-orbital for ${\rm Ge}$ atoms. We made this choice  analysing the composition of the density of states around the Fermi energy, as shown in Figs.~\ref{fig:fig2},~\ref{fig:fig3}, and~\ref{fig:fig4}.

We use \textsc{CRYSTAL14} to calculate the exchange coupling parameters to reduce the computational cost due to the size of the super-cell and therefore the high number of atoms. This code was especially useful for the calculation of the exchange couplings with the HSE06 hybrid functional. We have been able to switch from \textsc{QE} to \textsc{CRYSTAL14} after testing the consistency of the results with both codes, comparing electronic band structures and relaxed atomic positions. 
 
We use the \textsc{VESTA}~\cite{VESTA} and \textsc{Xcrysden}~\cite{Crysden} codes to visualize the geometrical structure and BZs, and to produce the plots in Fig.~\ref{fig:fig1}.

\subsection{Geometrical structure}
\label{sect:geometry}

\begin{figure*}[t]
\centering
\begin{overpic}[unit=1mm]{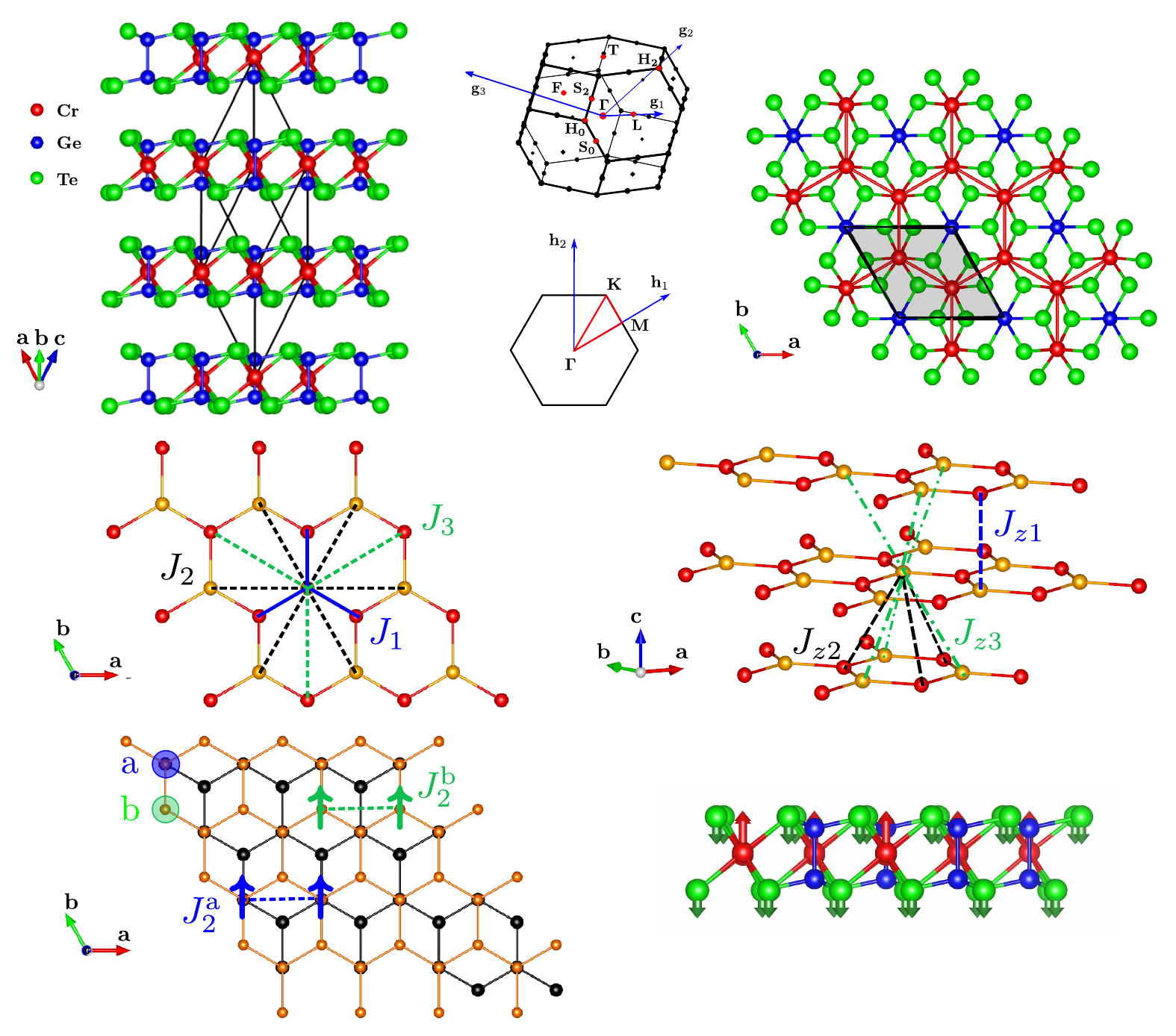} 
\put(10,140){\rm{a)}}
\put(63,140){\rm{b)}}
\put(100,140){\rm{c)}}
\put(63,100){\rm{d)}}
\put(10,75){\rm{e)}}
\put(82,75){\rm{f)}}
\put(10,35){\rm{g)}}
\put(82,35){\rm{h)}}
\end{overpic}
\caption{(Color online) a) Crystalline structure of bulk ${\rm Cr}_2{\rm Ge}_2{\rm Te}_6$. Different colors refer to different atomic species. In black, the rhombohedral unit cell. b) The corresponding Brillouin zone with the high-symmetry points used for the calculation of the electronic band structure\cite{Grosso2014,Hinuma2017}. 
 c) Top view of monolayer ${\rm Cr}_2{\rm Ge}_2{\rm Te}_6$. The shaded region denotes the unit cell. Notice the hexagonal sublattice formed by the Cr atoms (red). d) The corresponding Brillouin zone. e) Intra-layer exchange couplings between Cr atoms (orange and red refer to inequivalent atoms): $J_{1}$ denotes coupling between nearest-neighbor atoms;  $J_{2}$ denotes coupling between second-neighbor atoms; $J_{3}$ denotes coupling between third-neighbor atoms. f) Inter-layer exchange couplings: $J_{z1}$ (nearest-neighbor coupling), $J_{z2}$ (coupling between second-neighbor atoms), and $J_{z3}$  (coupling between third-neighbor atoms atoms). g) Top view of the Cr sublattices in AB-stacked bilayer Cr$_2$Ge$_2$Te$_6$. Orange: top layer. Black: bottom layer. As explained in the main text, in this case two different second-neighbor exchange couplings, denoted by $J^{\rm a}_{2}$ and $J^{\rm b}_{2}$, exist. h) Atomic distribution of the magnetic moments in the ferromagnetic phase. Note that the magnetic moment of the ${\rm Te}$ atom is smaller with respect to the one of the ${\rm Cr}$ atom, and oppositely oriented. The vectors are parallel to the easy magnetic axis---see Sect.~\ref{sect:Heisenberg}.  The magnetic moment of the ${\rm Ge}$ atom is not shown because it is too small compared to the magnetic moments of the other atoms. The length of the vectors is not to scale.
\label{fig:fig1}}
\end{figure*}
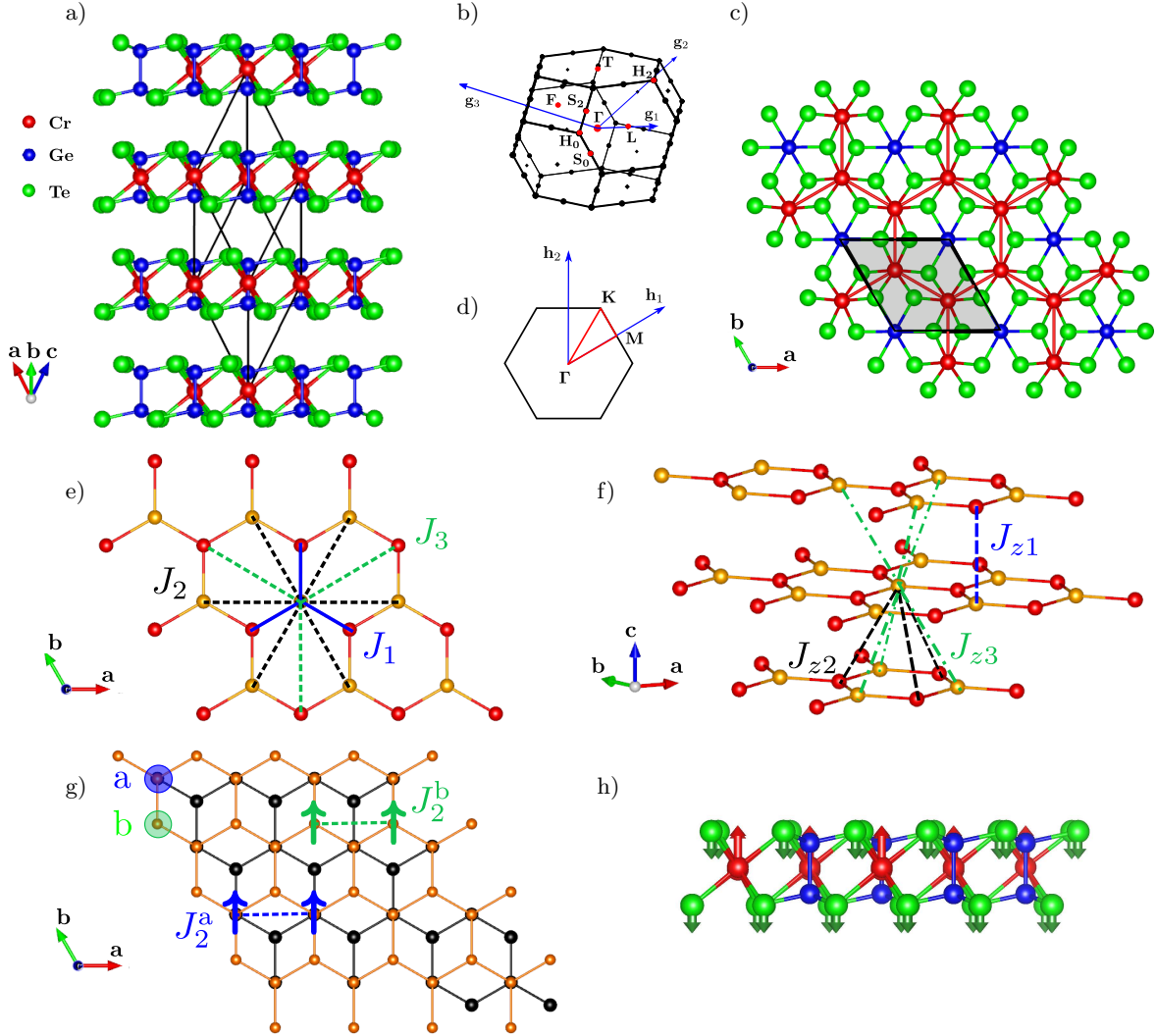

Bulk ${\rm Cr}_2{\rm Ge}_2{\rm Te}_6$ forms a layered structure with monolayers separated by a vdW gap---see Fig.~\ref{fig:fig1}a). Each monolayer---Fig.~\ref{fig:fig1}c)---is formed by edge-sharing ${\rm Cr}{\rm Te}_6$ octahedra where the ${\rm Ge}$ pairs are located in the hollow sites formed by the octahedra honeycomb. The layers are ABC-stacked, resulting in a rhombohedral $R\overline{3}$  symmetry (space group number 148), as shown in Fig.~\ref{fig:fig1}a), with experimental lattice constants~\cite{Carteaux1995,Siberchicot1996} $a=b=c=7.907$~\AA. This structure can also be described as ABC-stacked hexagonal crystal cells with experimental lattice constants $a'=b'=6.8275$~\AA~and $c'=20.5619$~\AA.  
 
Bilayer ${\rm Cr}_2{\rm Ge}_2{\rm Te}_6$ is formed by two AB-stacked monolayers---see Fig.~\ref{fig:fig1}g).
 
In Figs~\ref{fig:fig1}b) and~\ref{fig:fig1}d) we report the bulk and few-layer Brillouin Zones (BZs)~\cite{Grosso2014,Hinuma2017}, respectively.

We adopt the experimental rhombohedral unit cell for the calculation of the bulk properties, while for the monolayer and bilayer we use the hexagonal unit cell shown in  Fig.~\ref{fig:fig1}c). To simulate the monolayer (bilayer) forms, we consider a super-cell with about $15$~\AA~($22$~\AA) of vacuum along the $\hat{\bm z}$-direction between periodic images. 

We optimize the geometrical structures relaxing only the atomic positions until the components of all the forces on the ions are less than $10^{-3}~{\rm Ry}/{\rm Bohr}$, while we keep fixed the lattice parameters. Relaxation of the atomic positions has been carried out in the ferromagnetic phase. In Table~\ref{tab:tab1} we report the irriducible set of atomic coordinates and the main atomic distances and angles of both experimental~\cite{Carteaux1995,Siberchicot1996}  and optimized structures.

In order to disentangle the effect of the structure from that of magnetism, all calculations below use the atomic coordinates calculated with the spGGA. As can be seen in Table~\ref{tab:tab1}, the effect of nonlocal exchange interactions on the crystal structure is, anyway, small.

\begin{table*}[t]
 \caption{Irriducible set of atomic coordinates (Wyckoff positions in fractions of the crystallographic cell lattice vectors), main distances (in~\AA), and angles (in degrees) for bulk ${\rm Cr}_2{\rm Ge}_2{\rm Te}_6$. We present both the experimental values (column labelled by ``Exp.") and those  obtained by relaxing the atomic coordinates (columns labelled by ``spGGA" and ``HSE06", depending on the used approximation)---see Sect.~\ref{sect:geometry}. We did not find any appreciable differences between the internal coordinates of monolayer, bilayer, and bulk ${\rm Cr}_2{\rm Ge}_2{\rm Te}_6$.\label{tab:tab1}} 
{%
\newcommand{\mc}[3]{\multicolumn{#1}{#2}{#3}}
\begin{center}
\begin{tabular}{c|c|c|c||c|c|c||c|c|c}
&\mc{3}{c||}{Exp.\cite{Carteaux1995,Siberchicot1996}} & \mc{3}{c||}{spGGA}& \mc{3}{c}{HSE06}\\
\hline
& $x$ &$y$ & $z$ & $x$ & $y$ & $z$ & $x$ & $y$ & $z$ \\
\hline
Cr-18(f) & 0.0 & 0.0 & 0.3302 & 0.0 & 0.0 & 0.3343 & 0.0	& 0.0 &	0.3341  \\
\hline
Ge-6(c) & 0.0 & 0.0 & 0.059 & 0.0 & 0.0 & 0.0585 & 0.0	& 0.0 &	0.0577 \\
\hline
Te-6(c) & 0.663&	-0.033&	0.2482  & 0.6692 &	-0.0386 &	0.2492 & 0.6690 & 	-0.0334	& 0.2487  \\
\hline
Cr-Te~(\AA)	&		\mc{3}{c||}{2.749} &  \mc{3}{c||}{2.7629} &\mc{3}{c}{2.7802}\\
\hline
$\theta_{\rm Cr-Te-Cr}$~(deg)&    \mc{3}{c||}{91.6} & \mc{3}{c||}{91.081} &\mc{3}{c}{90.326}\\
\hline
Cr-Cr~(\AA)		&	\mc{3}{c||}{3.942} &  \mc{3}{c||}{3.942} &\mc{3}{c}{3.942} \\
\hline
Ge-Ge~(\AA)	&	\mc{3}{c||}{2.459} &  \mc{3}{c||}{2.4054} & \mc{3}{c}{2.3723}\\
\hline
Te-Ge~(\AA)	&		\mc{3}{c||}{2.578} & \mc{3}{c||}{2.6022} & \mc{3}{c}{2.5721} \\
\hline	
vdW gap~(\AA) &		\mc{3}{c||}{3.353} & \mc{3}{c||}{3.395} & \mc{3}{c}{3.374} \\
\hline
\end{tabular}
\end{center}
}
\end{table*}
\section{Electronic structure}
\label{sect:numerical_results}

In this Section we present our main results for the electronic band structures of  monolayer, bilayer, and bulk ${\rm Cr}_2{\rm Ge}_2{\rm Te}_6$.

\begin{figure*}[tp]
\centering
\begin{tabular}{cc}
  \begin{overpic}[unit=1mm,width=0.48\textwidth]{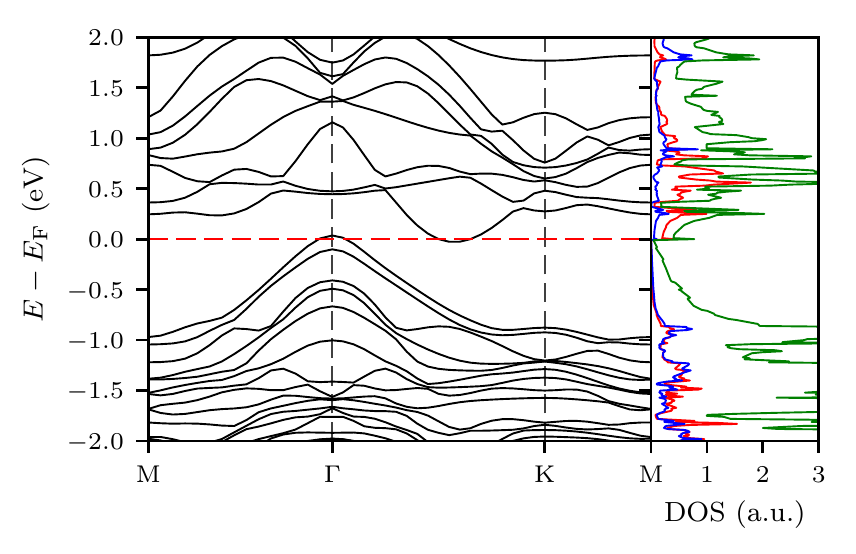} 
\put(3,50){\rm{a)}}
\end{overpic} &
\begin{overpic}[unit=1mm,width=0.48\textwidth]{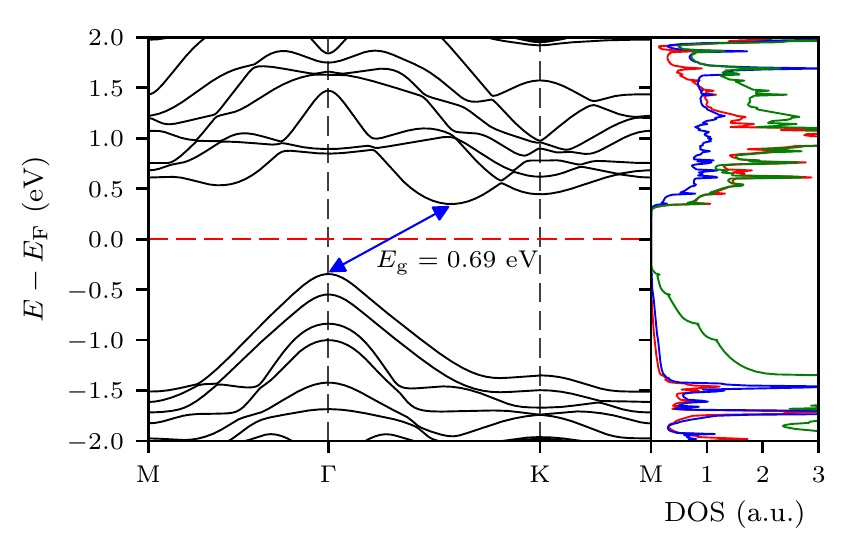}
\put(3,50){\rm{b)}}
\end{overpic}
\end{tabular}
 \caption{(Color online) Relativistic band structure and projected density of states (DOS) of {\it monolayer} ${\rm Cr}_2{\rm Ge}_2{\rm Te}_6$ in the ferromagnetic state. Panel a) Results obtained from the ${\rm GGA}+U$ approximation ($U=U_{\rm sc}=3.9~{\rm eV}$). Panel b) Results obtained from the HSE06 hybrid functional. In the DOS panels, colors refer to the DOS as projected onto the atomic orbitals of the various atoms (as in Fig.~\ref{fig:fig1}): Cr-$3d$ (red), Ge-$4p$ (blue), and Te-$5p$ (green). The HSE06 hybrid functional yields an insulating ferromagnetic state with an indirect gap $E_{\rm g} \sim 0.69~{\rm eV}$.
 \label{fig:fig2}}
\end{figure*}

\begin{figure*}[tp]
\centering
\begin{tabular}{cc}
  \begin{overpic}[unit=1mm,width=0.48\textwidth]{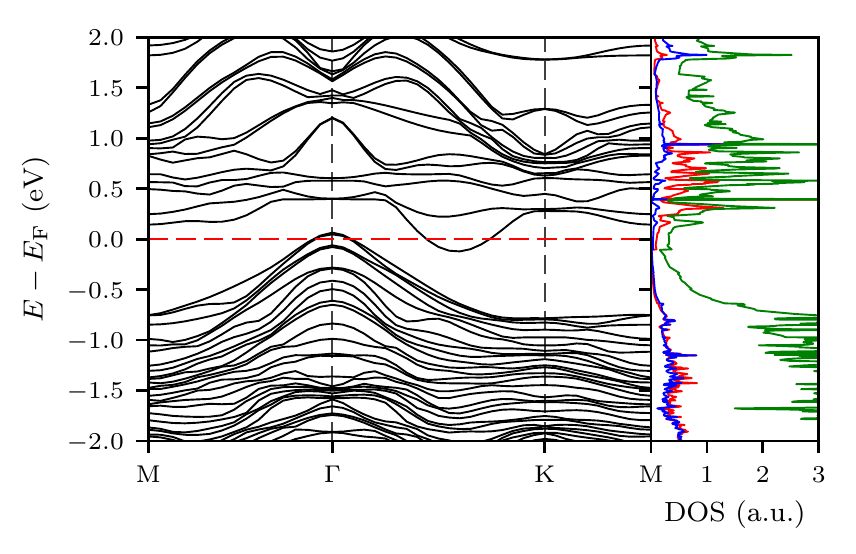} 
\put(3,50){\rm{a)}}
\end{overpic} &
\begin{overpic}[unit=1mm,width=0.48\textwidth]{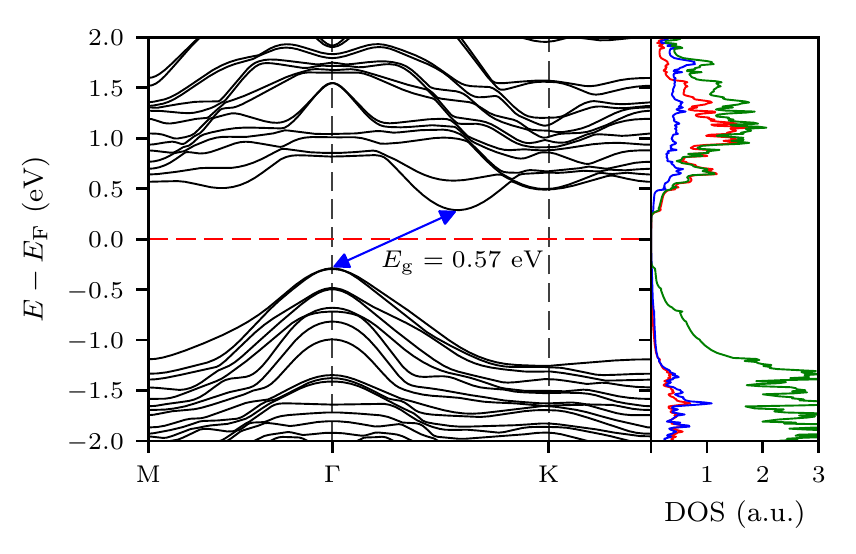}
\put(3,50){\rm{b)}}
\end{overpic}
\end{tabular}
 \caption{(Color online) Same as in Fig.~\ref{fig:fig2} but for {\it bilayer} Cr$_2$Ge$_2$Te$_6$. The HSE06 hybrid functional yields an insulating ferromagnetic state with an indirect gap of $0.57~{\rm eV}$.
 \label{fig:fig3}}
\end{figure*}

\begin{figure*}[tp]
\centering
\begin{tabular}{cc}
  \begin{overpic}[unit=1mm,width=0.48\textwidth]{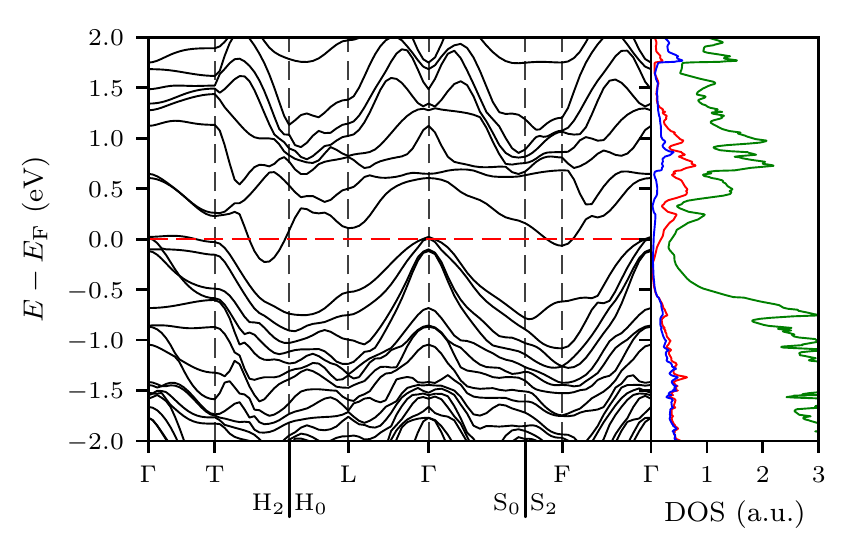} 
\put(3,50){\rm{a)}}
\end{overpic} &
\begin{overpic}[unit=1mm,width=0.48\textwidth]{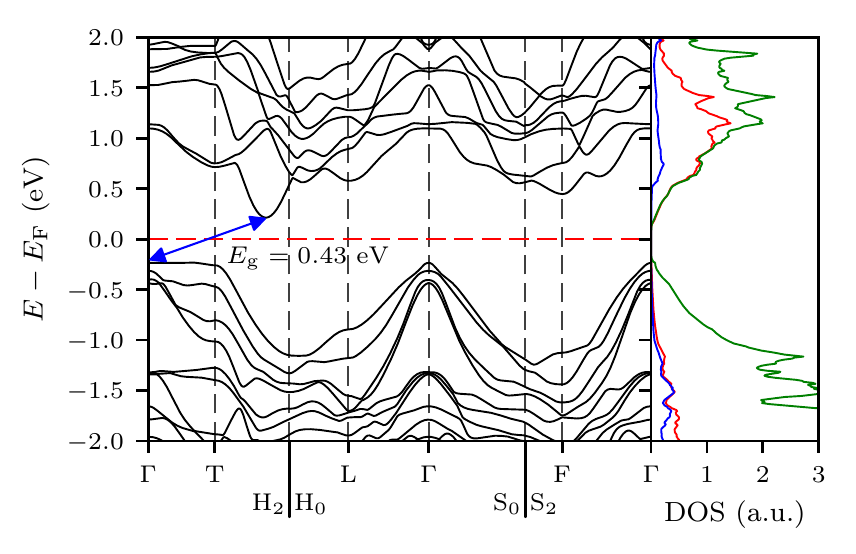}
\put(3,50){\rm{b)}}
\end{overpic}
\end{tabular}
 \caption{(Color online) Same as in Figs.~\ref{fig:fig2}-\ref{fig:fig3} but for {\it bulk} Cr$_2$Ge$_2$Te$_6$. The HSE06 hybrid functional yields an insulating ferromagnetic state with an indirect gap of $0.43~{\rm eV}$. This compares well with the only available experimental result~\cite{Li2018a}, $E^{\rm exp}_{\rm g} = 0.38~\rm{eV}$. 
\label{fig:fig4}}
\end{figure*}

In Figs.~\ref{fig:fig2},~\ref{fig:fig3}, and~\ref{fig:fig4}, we report the relativistic band structure of monolayer, bilayer, and bulk ${\rm Cr}_2{\rm Ge}_2{\rm Te}_6$, respectively. Results obtained in the ${\rm GGA}+U$ approximation are compared with those obtained with the HSE06 hybrid functional. As anticipated in Sect.~\ref{sect:technical_details}, the ${\rm GGA}+U$ results have been obtained by using the self-consistently-calculated value of the Hubbard-$U$ parameter, i.e.~$U_{\rm sc}=3.9~{\rm eV}$. 
The opposite qualitative behavior resulting from the two methods is evident.  Indeed, the ${\rm GGA}+U$ approximation---see Figs.~\ref{fig:fig2}a),~\ref{fig:fig3}a), and~\ref{fig:fig4}a)---leads to a metallic ground state. In stark contrast, the hybrid functional returns an insulating ground state with an indirect band gap of $E_{\rm g} \sim 0.69~{\rm eV}$  for monolayer ${\rm Cr}_2{\rm Ge}_2{\rm Te}_6$---see Fig.~\ref{fig:fig2}b)---$E_{\rm g} \sim 0.57~{\rm eV}$ for  bilayer ${\rm Cr}_2{\rm Ge}_2{\rm Te}_6$---see Fig.~\ref{fig:fig3}b)---and $E_{\rm g} \sim 0.43~{\rm eV}$ for bulk ${\rm Cr}_2{\rm Ge}_2{\rm Te}_6$---see Fig.~\ref{fig:fig4}b). The calculated electronic gap of the bulk crystal is comparable with recent experimental results based on angle-resolved photoemission spectroscopy~\cite{Li2018a}, which seem to indicate that bulk ${\rm Cr}_2{\rm Ge}_2{\rm Te}_6$ at $50~{\rm K}$ is a semiconductor with a gap $E^{\rm exp}_{\rm g}$ of at least $0.38~{\rm eV}$. A previous experiment~\cite{Ji2013} based on infrared transmission spectroscopy reported an optical gap of $\sim 0.74~{\rm eV}$ at $T=292~{\rm K}$. Since this study was carried out at $T\gg T_{\rm C}\sim 60~{\rm K}$, it does not help us in choosing the best xc functional to describe the ferromagnetic phase. To the best of our knowledge, Ref.~\onlinecite{Li2018a} is the only experimental guidance we have at the present stage, as optical studies (sensitive to the gap size) of few-layer ${\rm Cr}_2{\rm Ge}_2{\rm Te}_6$ in the ferromagnetic phase are currently unavailable.

Looking at the few-layer band structures in Figs.~\ref{fig:fig2} and~\ref{fig:fig3}, we observe that the valence band maximum (VBM) is at the $\Gamma$ point and the conduction band minimum (CBM) is located on the path that links the $\Gamma$ and ${\rm K}$ high-symmetry points. In the bulk case, instead, 
we see from Fig.~\ref{fig:fig4} that the CMB lies on the path that links the ${\rm T}$ and ${\rm H}_0$ high-symmetry points. The density of states (DOS) projected onto the atomic orbitals is also plotted in Figs.~\ref{fig:fig2}-\ref{fig:fig4}. Our results for the DOS demonstrate that the main contribution to the electronic bands near the Fermi energy comes from the ${\rm Cr}$ $3d$ and ${\rm Te}$ $5p$ orbitals, pointing out a strong hybridization between the ${\rm Cr}$ and ${\rm Te}$ orbitals. In fact, the ${\rm Te}$ atoms have a fundamental role in stabilizing the ferromagnetic phase of ${\rm Cr}_2{\rm Ge}_2{\rm Te}_6$ because they mediate super-exchange interactions in the ${\rm Cr}$-${\rm Te}$-${\rm Cr}$ bonds, as per the Goodenough-Kanamori rule~\cite{Goodenough1955,Kanamori1959}.
The DOS analysis shows also a small contribution from the ${\rm Ge}$ $4p$ orbitals near the Fermi energy.

\begin{figure*}
\begin{tabular}{cc}
\begin{overpic}[unit=1mm,width=0.48\textwidth]{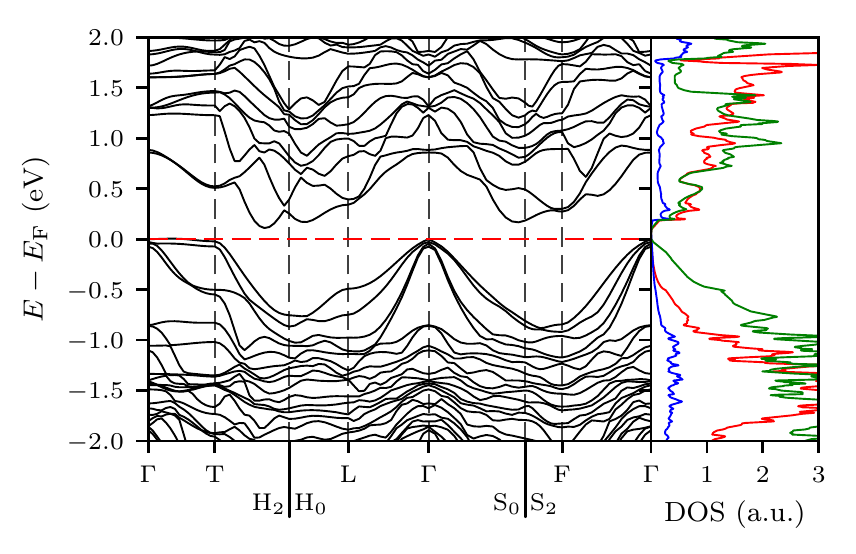}
\put(3,50){\rm{a)}}
\end{overpic} &
\begin{overpic}[unit=1mm,width=0.48\textwidth]{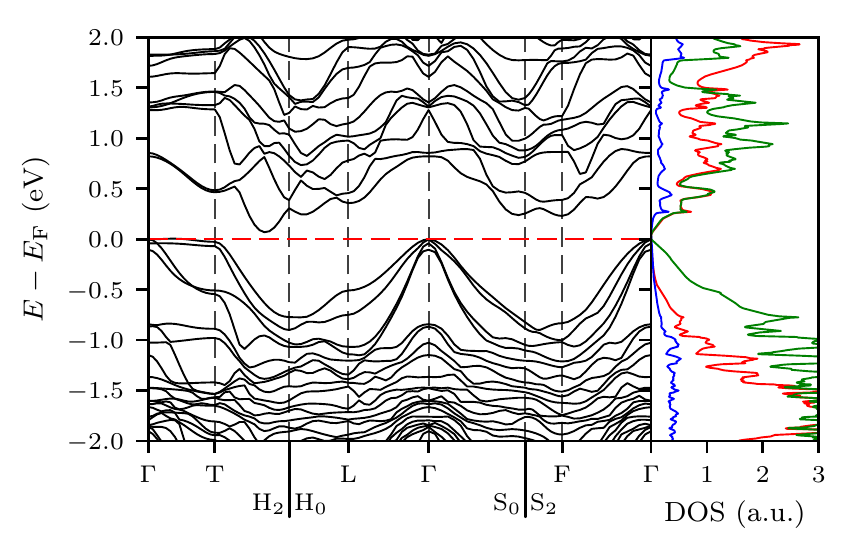}
\put(3,50){\rm{b)}}
\end{overpic} \\
\begin{overpic}[unit=1mm,width=0.48\textwidth]{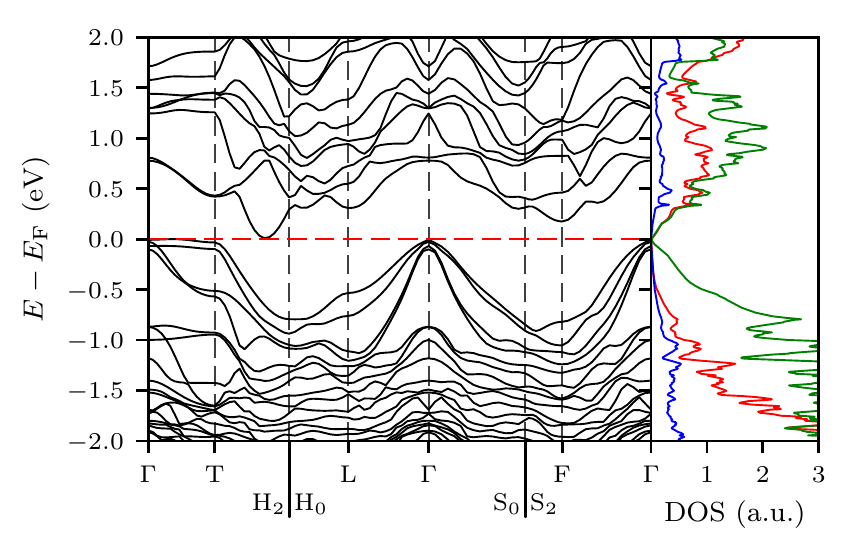}
\put(3,50){\rm{c)}}
\end{overpic}&
\begin{overpic}[unit=1mm,width=0.48\textwidth]{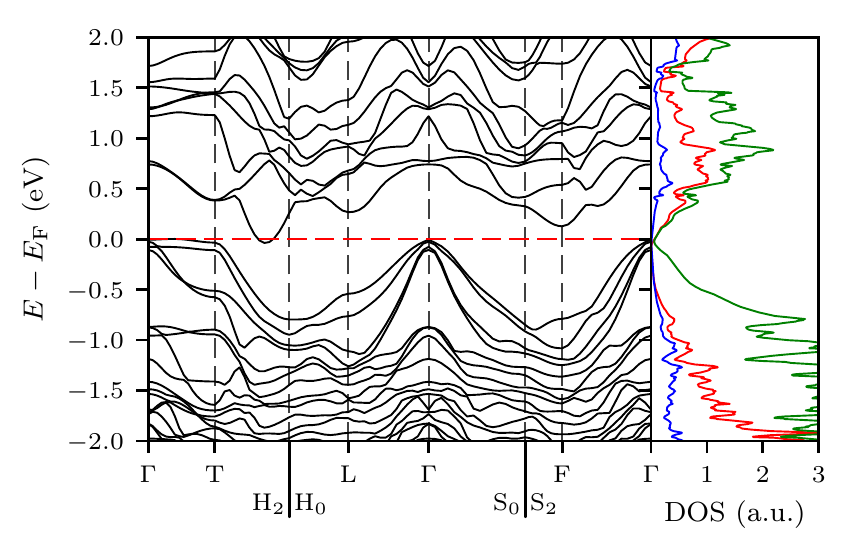}
\put(3,50){\rm{d)}}
\end{overpic}
\end{tabular}
\caption{(Color online) Relativistic band structure and projected density of states (DOS) of {\it bulk} Cr$_2$Ge$_2$Te$_6$ in the ferromagnetic state in the ${\rm GGA}+U$ approximation. Different panels refer to different values of the Hubbard-$U$ parameter: a) $U=0.0$~eV, b) $U=0.5$~eV, c) $U=1.0$~eV, and d) $U=1.5$~eV. In the DOS panels, colors refer to the DOS as projected onto the atomic orbitals of the various atoms (as in Figs.~\ref{fig:fig1}-\ref{fig:fig4}): Cr-$3d$ (red), Ge-$4p$ (blue), and Te-$5p$ (green). Note that the electronic band gap {\it decreases} with {\it increasing} $U$. The contribution of the Te-$5p$ orbitals does not change with $U$ while the contribution of the Cr-$3d$ orbitals decreases with increasing $U$ near the Fermi energy, due to a progressively weaker hybridization with the ${\rm Te}$-$5p$ orbitals.
\label{fig:fig5}}
\end{figure*}

In Fig.~\ref{fig:fig5} we show the relativistic band structure obtained in the ${\rm GGA}+U$ approximation, for different values of $0\leq U< U_{\rm sc}$. We note that the gap {\it decreases} with {\it increasing} $U$. We ascribe this behavior to the strong hybridization between the ${\rm Cr}$ and ${\rm Te}$ atoms, which leads to delocalized $e_{\rm g}$ states. These are not as sensitive as the other $d$ states from the Cr atoms  to the application of the Hubbard-$U$, due to their extended rather than atomic-like nature~\cite{Wu2019}. This is why we have used the HSE06 hybrid functional, which naturally takes into account nonlocal exchange interactions between all orbitals (and not only between the $3d$ orbitals of Cr atoms). By carrying out spin-polarized non-relativistic calculations, we have checked that HSE06 hybrid functional endows the VBM and CBM with spin-up character, differently from what happens in the ${\rm spGGA}$.
We conclude that the application of a  scissor operator to open the gap leads to wrong electronic and optical properties~\cite{Fang2018}. The calculated scalar-relativistic electronic gap for bulk ${\rm Cr}_2{\rm Ge}_2{\rm Te}_6$ is $E_{\rm g}\sim 0.73~{\rm eV}$, which is, as expected, larger than the one calculated with the inclusion of SOC. Finally, we note that the hybridization between the ${\rm Cr}$ $3d$ and ${\rm Te}$ $5p$ orbitals, which leads to the failure of the ${\rm GGA}+U$ approximation, is less evident in few-layer ${\rm Cr}_2{\rm Ge}_2{\rm Te}_6$ than in the bulk crystal. 

We now discuss about magnetism. With the HSE06 functional and in bulk ${\rm Cr}_2{\rm Ge}_2{\rm Te}_6$ the ferromagnetic state has lower energy than the paramagnetic one, the energy difference being of $\sim 3.5~{\rm eV}/{\rm Cr}$. Moreover, the ferromagnetic state is more stable than all the other antiferromagnetic configurations by at least $\sim 0.5~{\rm meV}/{\rm Cr}$ in the same approximation. 
In the ${\rm GGA}+U$ approximation, the magnetic properties of bilayer and bulk ${\rm Cr}_2{\rm Ge}_2{\rm Te}_6$, especially the interlayer exchange couplings, vary strongly with $U$. In fact, for small values of $U$, the ground state remains ferromagnetic, while for $U\gtrsim 1.5~{\rm eV}$ bilayer and bulk ${\rm Cr}_2{\rm Ge}_2{\rm Te}_6$ become antiferromagnetic, with oppositely-oriented magnetic moments in alternating layers~\cite{Kang2019}. This is in stark contrast with the experimental observations~\cite{Gong2017}. Monolayer ${\rm Cr}_2{\rm Ge}_2{\rm Te}_6$ remains in a ferromagnetic ground state for $1.5~{\rm eV}\lesssim U\leq U_{\rm sc}$, despite its exchange couplings greatly vary with $U$. In Fig.~\ref{fig:fig1}h) we report a pictorial representation of the magnetic moments, while in Table~\ref{tab:tab2} we summarize the values of the magnetic moments, as obtained by carrying out spin-polarized calculations---with the HSE06 functional and in the ${\rm GGA}+U_{\rm sc}$  approximation---for all the systems under study. The magnetic moments change slightly with the inclusion of SOC. As we will discuss in Sect.~\ref{sect:Heisenberg}, in  bilayer ${\rm Cr}_2{\rm Ge}_2{\rm Te}_6$ we observe two inequivalent ${\rm Cr}$ atoms in the unit cell, in one layer.

\begin{table}[bp]
 \caption{Atomic magnetic moments~(in units of the Bohr magneton $\mu_{\rm B}$) obtained with the HSE06 functional and the ${\rm GGA}+U$ approximation ($U=U_{\rm sc}=3.9~{\rm eV}$) for monolayer, bilayer, and bulk ${\rm Cr}_2{\rm Ge}_2{\rm Te}_6$.  Bilayer ${\rm Cr}_2{\rm Ge}_2{\rm Te}_6$ presents two inequivalent Cr atoms, ${\rm Cr}^{\rm a}$ and ${\rm Cr}^{\rm b}$---see Fig.~\ref{fig:fig1}g) and Sect.~\ref{sect:Heisenberg}.  The magnetic moments lie along the easy-axis.\label{tab:tab2}} 
{%
\newcommand{\mc}[3]{\multicolumn{#1}{#2}{#3}}
\begin{center}
\begin{tabular}{c|c|c|c}
\mc{2}{c|}{}                            & HSE06 & ${\rm GGA}+U_{\rm sc}$    \\
\hline\hline
\multirow{3}{*}{Monolayer} & ${\rm Cr}$ & 3.353  & 3.645                     \\\cline{2-4}
 & ${\rm Ge}$                           & 0.028  & 0.014                     \\\cline{2-4}
 & ${\rm Te}$                           & -0.121 & -0.167                    \\\cline{2-4}
\hline\hline
\multirow{4}{*}{Bilayer} & ${\rm Cr}^{\rm a}$ & 3.355  & 3.662                     \\\cline{2-4}
 & ${\rm Cr}^{\rm b}$                         & 3.361  & 3.667                     \\\cline{2-4}
 & ${\rm Ge}$                           & 0.029  & 0.013                     \\\cline{2-4}
 & ${\rm Te}$                           & -0.123 & -0.165                    \\\cline{2-4}
\hline\hline
\multirow{3}{*}{Bulk} & ${\rm Cr}$      & 3.378  & 3.698                     \\\cline{2-4}
 & ${\rm Ge}$                           & 0.027  & 0.013                     \\\cline{2-4}
 & ${\rm Te}$                           & -0.124 & -0.169                  
\end{tabular}
\end{center}
}
\end{table}

\section{Exchange couplings and magneto-crystalline anisotropy}
\label{sect:Heisenberg}

According to the Mermin-Wagner theorem~\cite{Mermin1966}, thermal fluctuations in 2D systems imply the impossibility of long-range magnetic order. A way to bypass the theorem and explain the discovery of 2D magnetic materials~\cite{Gong2017,Huang2017} is to recognize the presence of a large magneto-crystalline anisotropy. The latter lifts the invariance under rotations, an hypothesis of the Mermin-Wagner theorem. 

Electron spin resonance and ferromagnetic resonance measurements have been carried out in bulk ${\rm Cr}_2{\rm Ge}_2{\rm Te}_6$ over a wide frequency and temperature range~\cite{Zhang2016,Zeisner2018,Liu2019a,Khan2019}, revealing the development of 2D spin correlations in the vicinity of $T_{\rm C}$ and proving the intrinsically low-dimensional character of spin dynamics in this material. Such experiments returned a MAE of $\sim 50$-$90~\mu{\rm eV}/{\rm Cr}$, which is on the same order of what estimated in  Ref.~\onlinecite{Gong2017} for bulk ${\rm Cr}_2{\rm Ge}_2{\rm Te}_6$.

In order to understand the microscopic properties of magnetic materials it is crucial to derive an effective spin model. In the case of localized magnetic moments, the Heisenberg spin Hamiltonian can be used. The total Hamiltonian is the sum a non-magnetic part and a magnetic part, which has the form 
\begin{equation}\label{eq:Heisenberg+MAE}
\hat{\cal H}_{\rm spin}=\sum_{i<j} J_{ij}\hat{\bm S}_i\cdot \hat{\bm S}_j+\sum_{i}A_i\hat{S}^2_{iz}~.
\end{equation}
The first term on the right-hand side of Eq.~(\ref{eq:Heisenberg+MAE}) is the rotationally-invariant Heisenberg Hamiltonian with exchange couplings $J_{ij}$ and the sums over $i<j$ run over all the ${\rm Cr}$ pairs without double counting. The second term is the magneto-crystalline anisotropy energy (MAE), with the sum over $i$ running over all the ${\rm Cr}$ sites. The Dzyaloshinski-Moriya interaction term, $\sum_{i<j}{\bm D}_{ij}\cdot(\hat{\bm S}_i\times\hat{\bm S}_j)$, is absent in Eq.~(\ref{eq:Heisenberg+MAE}) due to the existence of an inversion center between the ${\rm Cr}$ ions~\cite{Moriya1960}. 

The MAE originates from SOC~\cite{Bruno1989,Li2014}, and is therefore a quantum effect of relativistic nature. It breaks the rotational invariance with respect to the spin quantization axis and is determined by the interaction between the orbital state of a magnetic ion and the surrounding crystalline field. It can be calculated~\cite{Gong2017,Li2018,Li2014} by looking at the total energy difference---obtained through self-consistent calculations in the presence of SOC---between the configuration with all spins perpendicular to the layer (along the $\hat{\bm z}$ direction) and that with all spins parallel to the layer (along the $\hat{\bm x}$ or $\hat{\bm y}$ directions).  A positive (negative) sign of the MAE means that the system is an easy-plane (easy-axis) ferromagnet.  

As anticipated in Sect.~\ref{sect:intro}, an accurate method to calculate the exchange couplings for small-gap semiconductors and metals is the FSMA approach~\cite{Whangbo2003,Xiang2011,Xiang2013a}. The FSMA approach considers one specific magnetic pair at a time in a super-cell. Without loss of generality, the energy can be written as
\begin{equation}\label{eq:energies}
\begin{split}
 E=&E_{\rm spin}+E_{0}=\\
 =&J_{12}\bm{S}_1\cdot\bm{S}_2+\bm{S}_1\cdot\bm{K}_1+\bm{S}_2\cdot\bm{K}_2+E_{\rm others}+E_{0}~,
 \end{split}
\end{equation}
where $E_{\rm spin}$ and $E_{0}$ represent the magnetic and non-magnetic energies, respectively. Here, $J_{12}\bm{S}_1\cdot\bm{S}_2$ is the exchange coupling between sites $1$ and $2$, while $\bm{S}_i\cdot\bm{K}_i$ represents the coupling between site $i$ and the all the magnetic sites different from $1$ and $2$. Finally, $E_{\rm others}$ represents the contribution to the energy stemming from the interaction between sites $1$ and $2$ and all the other non-magnetic sites. Consider the following collinear spin states:
\begin{center}
\begin{tabular}{ccc}
State & $\bm{S}_1$ & $\bm{S}_2$\\
1 & (0,0,S) & (0,0,S)\\
2 & (0,0,S) & (0,0,-S)\\
3 & (0,0,-S) & (0,0,S)\\
4 & (0,0,-S) & (0,0,-S)
\end{tabular}
\end{center}
The other spin states are kept fixed, according to the ground-state spin configuration.
Using Eq.~(\ref{eq:energies}), we can write the energies of the four states as following:
\begin{align*}
 &E_1 = E_0 + J_{12}S^2+\bm{S}_1\cdot\bm{K}_1+\bm{S}_2\cdot\bm{K}_2~, \\
 &E_2 = E_0 - J_{12}S^2+\bm{S}_1\cdot\bm{K}_1-\bm{S}_2\cdot\bm{K}_2~,  \\
 &E_3 = E_0 - J_{12}S^2-\bm{S}_1\cdot\bm{K}_1+\bm{S}_2\cdot\bm{K}_2~,  \\
 &E_4 = E_0 + J_{12}S^2+\bm{S}_1\cdot\bm{K}_1-\bm{S}_2\cdot\bm{K}_2~.
\end{align*}
The solution of this system leads to the equality
\begin{equation}
 J_{12}=\dfrac{E_1-E_2-E_3+E_4}{4S^2}~.
\end{equation}

\begin{figure}[t]
\begin{tabular}{c}
\begin{overpic}[width=\columnwidth]{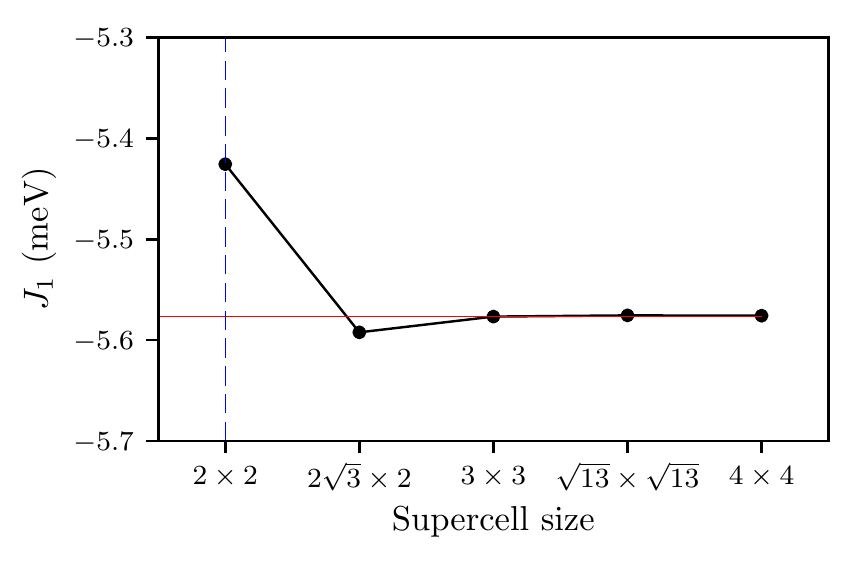}
\put(1,150){\rm{a)}}
\end{overpic} \\
\begin{overpic}[width=\columnwidth]{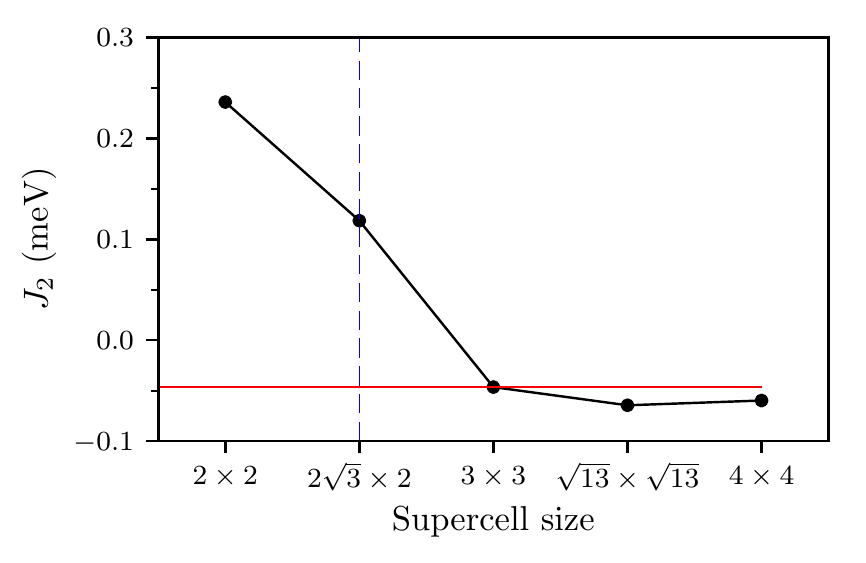}
\put(1,150){\rm{b)}}
\end{overpic} \\
\begin{overpic}[width=\columnwidth]{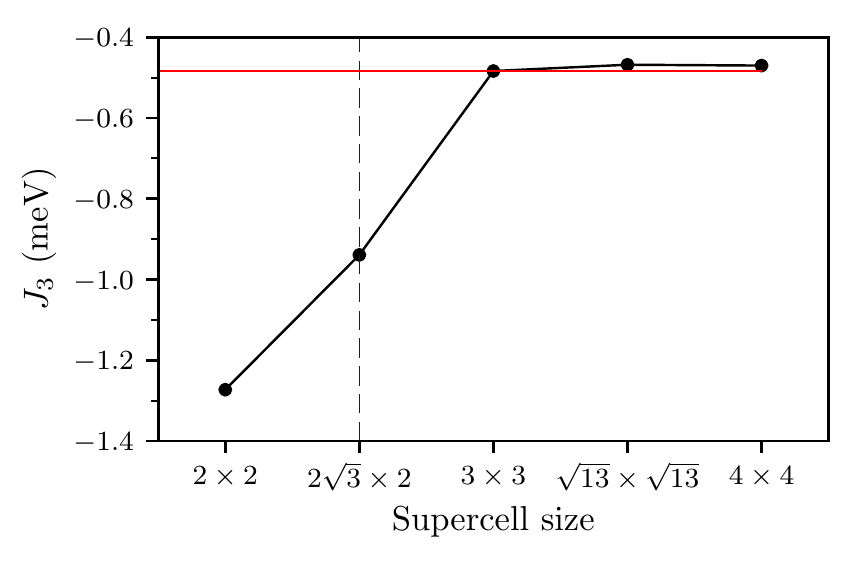}
\put(1,150){\rm{c)}}
\end{overpic}
\end{tabular}
\caption{(Color online) Convergence of the intra-layer exchange coupling parameters $J_i$ with the size of the super-cell. Results in this figure refer to monolayer ${\rm Cr}_2{\rm Ge}_2{\rm Te}_6$ and have been obtained in the spGGA. The horizontal thin lines denote the converged values reported in Table~\ref{tab:tab3} (spGGA). The vertical dashed lines denote the values of the super-cell size used in the literature~\cite{Xu2018,Gong2017,Li2018}.
\label{fig:fig6}}
\end{figure}

The FSMA method is accurate when the super-cell used for the calculations is large enough. We monitor the accuracy by using a convergence test with respect to the super-cell size. In Fig.~\ref{fig:fig6} we show the convergence of the intra-layer exchange coupling parameters $J_i$, depicted in Fig.~\ref{fig:fig1}e), with the size of the super-cell for monolayer ${\rm Cr}_2{\rm Ge}_2{\rm Te}_6$. We clearly see that we need at least a $3\times 3$ ($3\times 3 \times 1$) supercell for monolayer (bulk) ${\rm Cr}_2{\rm Ge}_2{\rm Te}_6$ to have  converged results. This implies that the results reported in Refs.~\onlinecite{Gong2017,Xu2018,Li2018} for the exchange couplings are not converged. In particular, we see that the coupling $J_2$ between second neighbors changes character from anti-ferromagnetic to ferromagnetic if we choose a $3\times 3$ supercell instead of a $2\times 2$ or a $2\sqrt{3}\times 2$ supercell, see Fig.~\ref{fig:fig6}b).

In Table~\ref{tab:tab3} we present a summary of our results (in black) and the comparison with existing calculations (in red). All the exchange couplings have been calculated with the FSMA approach. The only exception is the row in Table~\ref{tab:tab3} labeled by ``${\rm spGGA}^{\prime}$''. Results in this row have been calculated with the FM-AFM method (see Sect.~\ref{sect:intro}) with the only aim of pointing out how much these differ from the ones calculated with the FSMA, in the same approximation (spGGA) for the xc functional. Note that, for monolayer and bilayer ${\rm Cr}_2{\rm Ge}_2{\rm Te}_6$, the sign of the MAE calculated in the spGGA (${\rm MAE}>0$) is opposite to that calculated with the HSE06 hybrid functional (${\rm MAE}<0$). For all the calculations and the comparisons with the literature we use $S = 3/2$, as justified by the atomic magnetic moments reported in Table~\ref{tab:tab2}. 

Following Ref.~\onlinecite{Torelli2018} and taking into account only the largest exchange coupling (i.e.~$J_1$) and the MAE, we estimate the Curie temperature $T_{\rm C}$. We find: $T_{\rm C}\sim 34~{\rm K}$ for monolayer ${\rm Cr}_2{\rm Ge}_2{\rm Te}_6$; $T_{\rm C}\sim 40~{\rm K}$ for bilayer ${\rm Cr}_2{\rm Ge}_2{\rm Te}_6$; and $T_{\rm C}\sim 48~{\rm K}$ for bulk ${\rm Cr}_2{\rm Ge}_2{\rm Te}_6$. The trend of $T_{\rm C}$ with respect to the number of layer is in qualitative agreement with the experimental results of Ref.~\onlinecite{Gong2017}. The  results for $J_i$, MAE, and $T_{\rm C}$ emphasize a strong dimensionality effect, in agreement with the experimental behaviors~\cite{Gong2017}.

We finally note that bilayer ${\rm Cr}_2{\rm Ge}_2{\rm Te}_6$ presents a particular magnetic structure resulting from the lack of the ABC-stacking configuration and periodicity along the $\hat{\bm z}$ direction. Indeed, in  AB-stacked bilayer ${\rm Cr}_2{\rm Ge}_2{\rm Te}_6$ we can identify two inequivalent Cr atoms in the unit cell---see Fig.~\ref{fig:fig1}g). With reference to this figure, we clearly see that in the top layer (orange) there are Cr atoms which have another Cr atom underneath in the bottom layer (black)---these Cr atoms are label by the letter ``a"  in blue in Fig.~\ref{fig:fig1}g)---while there are other Cr atoms, which do not have a Cr partner underneath in the bottom layer---these Cr atoms are label by the letter ``b"  in green in Fig.~\ref{fig:fig1}g). In Table~\ref{tab:tab3} we report the calculated values of the two associated second-neighbor exchange couplings, $J^{\rm a}_2\neq J^{\rm b}_2$. Note that the inter-layer couplings affect the interaction between these two inequivalent Cr atoms. This is evident if we compare the results obtained with the spGGA and HSE06 functionals. In fact, the spGGA overestimates the inter-layer couplings, resulting in a vanishing $J^{\rm b}_2$. In contrast, smaller inter-layer couplings, extracted from the HSE06 functional, lead to two different but comparable second-neighbor exchange couplings $J^{\rm a}_2$ and $J^{\rm b}_2$. In the bulk crystal the equivalence between the two Cr atoms is restored and we find $J^{\rm a}_2=J^{\rm b}_2= J_2$.
\begin{table*}[bp]
 \caption{(Color online) Exchange couplings $J_i$ and magneto-crystalline anisotropy energy (MAE)  for monolayer, bilayer, and bulk ${\rm Cr}_2{\rm Ge}_2{\rm Te}_6$. In black: results obtained in this work. Different rows refer to different approximations (see main text). In red: results available in the cited literature. (If a cell is empty it means that the corresponding result was not found in the literature.) All the exchange couplings have been calculated with the FSMA approach (see Sect.~\ref{sect:Heisenberg}). The only exception is the row labeled by ``${\rm spGGA}^{\prime}$'', which presents results calculated with the FM-AFM method (see Sect.~\ref{sect:intro}).\label{tab:tab3}} 
{%
\newcommand{\mc}[3]{\multicolumn{#1}{#2}{#3}}
\begin{center}
\begin{tabular}{c|c|c|c|c|c|c|c}
\mc{7}{c}{Monolayer}\\
\hline
\hline 
 & \mc{2}{c|}{$J_1 ({\rm meV})$}& \mc{2}{c|}{$J_2 ({\rm meV})$}& \mc{2}{c|}{$J_3 ({\rm meV})$}& MAE~($\mu\rm{eV}/{\rm Cr}$)\\
 \hline
{\rm spGGA} & \mc{2}{c|}{-5.577}& \mc{2}{c|}{-0.046} & \mc{2}{c|}{-0.484}   &62\\
\hline 
${\rm spGGA}^{\prime}$&\mc{2}{c|}{-10.485} & \mc{2}{c|}{0.344} & \mc{2}{c|}{-0.678}    &62\\
\hline
HSE06& \mc{2}{c|}{-6.236} & \mc{2}{c|}{-0.083} & \mc{2}{c|}{-0.268}  &-80\\ 
\hline
\color{red}Xu2018\cite{Xu2018}& \mc{2}{c|}{\color{red}-6.64} & \mc{2}{c|}{} & \mc{2}{c|}{}  & \color{red} 250\\ 
\hline
\color{red}Li2014\cite{Li2014b}& \mc{2}{c|}{\color{red}-5.18} & \mc{2}{c|}{} & \mc{2}{c|}{}  & \\ 
\hline
\color{red}Sivadas2015\cite{Sivadas2015}& \mc{2}{c|}{\color{red}-3.68} & \mc{2}{c|}{\color{red}0.39} & \mc{2}{c|}{\color{red}0.43}  &  \\ 
\hline
\color{red}Zhuang2015\cite{Zhuang2015}& \mc{2}{c|}{\color{red}-3.07} & \mc{2}{c|}{} & \mc{2}{c|}{}  & \color{red} -220 \\ 
\hline
\color{red}Fang2018\cite{Fang2018}& \mc{2}{c|}{\color{red}-5.662} & \mc{2}{c|}{\color{red}0.040} & \mc{2}{c|}{\color{red}-0.129}  & \color{red} -107 \\ 
\hline
\color{red}Dong2019\cite{Dong2019}& \mc{2}{c|}{\color{red}-6.7} & \mc{2}{c|}{} & \mc{2}{c|}{}  & \color{red} -320 \\ 
\hline
\mc{7}{c}{Bilayer}\\
\hline
\hline 
 & $J_1 ({\rm meV})$& $J^{\rm a}_2/J^{\rm b}_2 ({\rm meV})$& $J_3 ({\rm meV})$& $J_{z1} ({\rm meV})$& $J_{z2} ({\rm meV})$& $J_{z3} ({\rm meV})$ & MAE~($\mu$\rm{eV}/~${\rm Cr}$)\\
\hline
spGGA & -5.694 & -0.0429/-0.001 & -0.629 & -0.096 & -0.024 & -0.365&40\\
\hline
HSE06 &  -6.399 & -0.096/-0.076 & -0.432 & -0.017 & 0.196 & -0.199&-140\\
\hline
\color{red}Fang2018\cite{Fang2018} & \color{red}-6.338 &\color{red} 0.209 & \color{red}-0.475 &\color{red} -0.177 &\color{red}  &\color{red} &\color{red}-274\\
\hline
\mc{7}{c}{Bulk}\\
\hline
\hline 
 & $J_1 ({\rm meV})$& $J_2 ({\rm meV})$& $J_3 ({\rm meV})$& $J_{z1} ({\rm meV})$& $J_{z2} ({\rm meV})$& $J_{z3} ({\rm meV})$ & MAE~($\mu$\rm{eV}/~${\rm Cr}$)\\
\hline
spGGA & -5.755 & -0.028 & -0.450 & -0.1097 & -0.043 & -0.359&-80\\
\hline
HSE06 & -6.607 & -0.083 & -0.627 & -0.010 & 0.190 & -0.225&-260\\
\hline
\color{red}Gong2017\cite{Gong2017,Li2018} & \color{red}-3.76 &\color{red} 0.08 & \color{red}-0.16 &\color{red} 0.05 &\color{red} -0.12 &\color{red} -0.38&\color{red}-50\\
\hline
\color{red}Fang2018 \cite{Fang2018} & \color{red}-6.995 &\color{red} 0.346 & \color{red}-0.773 &\color{red} -0.480 &\color{red}  &\color{red} &\color{red}-471\\
\hline
\end{tabular}
\end{center}
}
\end{table*}
\section{Summary and conclusions}
\label{sect:conclusions}

In summary, in this Article we have reported on a comparative density functional theory study of the electronic and magnetic properties of the recently discovered atomically-thin magnetic material ${\rm Cr}_2{\rm Ge}_2{\rm Te}_6$. 

We have clearly shown that the often-used ${\rm LDA}+U$ and ${\rm GGA}+U$ approaches fail in predicting the ground-state properties of this material in both its monolayer and bilayer forms, and even more spectacularly in its bulk form. According to these approaches, ${\rm Cr}_2{\rm Ge}_2{\rm Te}_6$ should have a metallic ground state, with its magnetic properties strongly depending on the number of layers and the values of the Hubbard-$U$ parameter used, including the self-consistently-calculated one, $U_{\rm sc}=3.9~{\rm eV}$ for the bulk crystal. On the contrary, the use of the HSE06 hybrid functional, which naturally takes into account nonlocal exchange interactions between all orbitals (and not only between the $d$ orbitals of Cr), yields an insulating ferromagnetic ground state---see Figs.~\ref{fig:fig2}-\ref{fig:fig4}---and an indirect gap, which, for the case of bulk ${\rm Cr}_2{\rm Ge}_2{\rm Te}_6$, is comparable with the one recently measured~\cite{Li2018a} via angle-resolved photo-emission spectroscopy.

Finally, we have calculated all the relevant exchange couplings for monolayer, bilayer, and bulk ${\rm Cr}_2{\rm Ge}_2{\rm Te}_6$, comparing results obtained via different methods and/or super-cell size---see Table~\ref{tab:tab3}. Using the four-state mapping analysis approach~\cite{Whangbo2003,Xiang2011,Xiang2013a}, we have shown that the slow convergence of the exchange couplings with the super-cell size may be responsible for discrepancies in the literature~\cite{Gong2017,Li2018,Xu2018}. Such a convergence check has been reported in Fig.~\ref{fig:fig6}. 

In the case of bilayer ${\rm Cr}_2{\rm Ge}_2{\rm Te}_6$, we have shown that two distinct intra-layer second-neighbor exchange couplings emerge---see the portion of Table~\ref{tab:tab3} that concerns bilayer ${\rm Cr}_2{\rm Ge}_2{\rm Te}_6$.

In the future, it would be interesting to study the dependence of these results on doping, obtained via the usual electrical field effect, plasmons, and plasmon-magnon coupling in ${\rm Cr}_2{\rm Ge}_2{\rm Te}_6$ and other magnetic 2D materials.

\acknowledgments
This work was supported by the European Union's Horizon 2020 research and innovation programme under grant agreement No. 785219 - GrapheneCore2. G.M. was supported by the EDISON-Volta prize (2018). We thank L. Paulatto for help with the \textsc{open\_grid} package of \textsc{Quantum Espresso} and C. Tresca and M.I. Katsnelson for useful discussions. Finally, we acknowledge the allocation of computer resources from CINECA, through the ``ISCRA C'' projects ``HP10CXQECI'' and ``HP10CXRX1O'', PRACE (Project No. 2017174186), and EDARI (Grant A0050901202).

\end{document}